\documentclass[11pt,oneside]{amsart}

\usepackage{amsmath,amssymb,bm,psfrag}
\usepackage[dvips]{graphicx}

\theoremstyle{plain} 
\newtheorem{theorem}{Theorem}[section]
\newtheorem{lemma}[theorem]{Lemma}
\newtheorem{corollary}[theorem]{Corollary}

\newtheorem{remark}[theorem]{Remark}

\numberwithin{equation}{section}

\begin{document}

\begin{center}

{\bf\large Ancestral Graph with Bias in Gene Conversion}

\vspace{1cm}

Shuhei Mano

{\it The Institute of Statistical Mathematics,\\
Tachikawa 190-8562, Japan\\
and\\
Japan Science and Technology Agency,\\
Kawaguchi 332-0012, Japan}

\end{center}

\vspace{3cm}

\noindent
Address for correspondence: The Institute of Statistical Mathematics, 
10-3 Midori-cho, Tachikawa, Tokyo 190-8562, Japan; Email: {\tt smano@ism.ac.jp}

\newpage

\noindent
{\bf Abstract}

\noindent
Gene conversion is a mechanism by which a double-strand break in a DNA 
molecule is repaired using a homologous DNA molecule as a template. As 
a result, one gene is 'copied and pasted' onto the other gene. It was 
recently reported that the direction of gene conversion appears to be 
biased towards G and C nucleotides. In this paper a stochastic model of 
the dynamics of the bias in gene conversion is developed for a finite 
population of members in a multigene family. The dual process is 
the biased voter model, which generates an ancestral random graph for 
a given sample. An importance-sampling algorithm for computing 
the likelihood of the sample is also given. 

\vspace{1cm}


\newpage

\section{Introduction} 

Gene conversion is a mechanism by which a double-strand break in a DNA 
molecule is repaired using a homologous DNA molecule as a template. As 
a result, the homologous DNA fragments become identical (i.e., one gene 
is 'copied and pasted' onto the other gene). In diploid organisms, gene 
conversion can occur between orthologous DNA molecules of paired 
homologous chromosomes during recombination, referred to as {\em allelic 
gene conversion}. Alternatively, {\em ectopic gene conversion} can occur 
among paralogous DNA molecules of duplicated gene copies in different 
loci, called a multigene family.

Evolutionary mechanisms are not rigorously tuned. The direction of the 
conversion appears biased towards G and C. According to this hypothesis, 
when an AT versus GC polymorphism exists in homologous DNA molecules, 
the A or T variant is more likely to be converted to G or C than the 
reverse. Regions of a genome that evolve rapidly have been regarded as 
being under strong positive selection. Surprisingly enough, it was 
reported that many protein coding changes in the fastest evolving genes 
of the human genome are not a result of positive selection but a result 
of biased fixation of AT to GC mutations~\cite{Berglund2009}. 
In the histone paralogous genes of humans and mice, gene copies that 
belong to subfamilies with very similar sequences, which are presumably 
undergoing ectopic gene conversion, have higher GC content than unique 
gene copies, which are free from ectopic gene conversion~\cite{Galtier2003}. 
The result of the bias in allelic gene conversion is indistinguishable 
from that of natural selection, since the models are mathematically 
identical to each other~\cite{Nagylaki1983}. In contrast, the dynamics 
of ectopic gene conversion are poorly understood, although some theoretical 
models have been developed~\cite{NagylakiPetes1982,Walsh1985}.

In this paper a stochastic model of the dynamics of bias in ectopic gene 
conversion in a finite population is developed. A corresponding island model 
of population subdivision~\cite{Wright1951} with allele-dependent migration, 
with a diffusive limit identical to that of the model of ectopic gene 
conversion, is introduced. The model is formulated in terms of a biased voter 
model~\cite{Harris1976} on complete graphs, where the interactions between 
demes are complete and each deme is a complete graph of sites. The biased 
voter model has a dual process and the limit process generates a random 
graph for a given sample that is analogous to the coalescent 
genealogy~\cite{Kingman1982}. We call the graph the ancestral bias graph. 
The ancestral bias graph is similar to the ancestral selection graph, which 
was introduced by \cite{KroneNeuhauser1997}, but the ancestral bias graph 
is structured with allele dependent migration. An importance-sampling 
algorithm that can be used to compute the likelihood of a given sample 
is provided. The algorithm is applied to the mouse histone H2A gene family 
data set.

\section{The Model}

Consider a monoecious panmictic population that consists of $N$ 
haploid individuals, who have a size $d\,(\ge 2)$-unlinked multigene 
family (i.e., duplicated gene copies at unlinked $d$ loci on 
distinct chromosomes). Assume the population evolves according to 
a continuous-time Moran model, in which an individual produces one 
offspring at a time. The type of the offspring is modified from that
of the parent according to mutation and gene conversion mechanisms.
The offspring will then replaces an individual chosen at random from 
the population. The offspring may replace its own parent. 
The replaced individual is removed from the population, keeping 
the population size constant. We assume that an individual reproduces 
at a rate of $\lambda_N$.

Assume in the multigene family there are two types of genes:
allele $A$ and allele $a$. Let $c$ be the rate at which a gene at 
a particular locus in an offspring is converted by a gene of any one 
of the other $d-1$ loci in the offspring with equal probability. 
Only a subset of the total conversion events involves different alleles. 
Among such conversion events involving different alleles, let 
$(1+b)/2$ be the fraction of these events that results in an allele $a$ 
being converted by an allele $A$, and similarly, let $(1-b)/2$ be 
the fraction of the events that result in an allele  $A$ being converted 
by an allele $a$, where $0\le b\le 1$. The conversion event is biased if 
$b>0$~\cite{NagylakiPetes1982}. The rate at which an allele $a$ is 
converted by an allele $A$, and an allele $A$ is converted by an allele 
$a$ is $c(1+b)/(d-1)$ and $c(1-b)/(d-1)$, respectively, where 
$c\, (0<c<1)$ is the conversion rate. For example, when $d=3$, 
an individual of type $AAa$ produces an offspring of type $Aaa$, $aAa$, 
and $AAA$ at rates $c(1-b)/2$, $c(1-b)/2$, and $c(1+b)$, respectively. 
For each locus, an offspring will have the same allelic type 
as the parent with a probability of $1-u$ and will have the other type
with a probability of $u$. The coincidence of a gene conversion or 
a mutation is ignored.

The state of the population at time $t$ can be represented as 
a continuous-time Markov chain ${\mathbf W}^N(t)=(W^N_\alpha(t))$, 
where $W^N_\alpha(t)$ is the number of individuals of type 
$\alpha \in \{A,a\}^d$ in the population at time $t$. If 
${\mathbf W}^N(t)={\mathbf w}$, the transition to 
${\mathbf w}+{\mathbf e}_\alpha$ occurs at a rate of
\begin{equation}
\lambda_N w_{\alpha}\frac{N-w_{\alpha}}{N}
(1-\sum_{\beta\neq\alpha}q_{\alpha\beta})
+\lambda_N\frac{N-w_{\alpha}}{N}\sum_{\beta\neq\alpha}w_{\beta}
q_{\beta\alpha}
\label{trans_multi_1}
\end{equation}
and the transition to ${\mathbf w}-{\mathbf e}_\alpha$ occurs at a rate of
\begin{equation}
\lambda_N w_{\alpha}\frac{w_{\alpha}}{N}
\sum_{\beta\neq\alpha}q_{\alpha\beta}
+\lambda_N\frac{w_{\alpha}}{N}\sum_{\beta\neq\alpha}w_{\beta}
(1-q_{\beta\alpha}),
\label{trans_multi_2}
\end{equation}
where $q_{\alpha\beta}$ is the rate at which an individual of type
$\alpha$ changes to type $\beta$. For example, 
$q_{AAa,Aaa}=c(1-b)/2+u(1-u)^2$. $X_i^N(t)$ denotes the fraction of 
allele $a$ in the $i$-th locus at time $t$. The limiting diffusion 
approximation is obtained by setting $\lambda_N=N/2$ assuming that 
$Nu\rightarrow\theta$ and $Nc\rightarrow \gamma$ as $N\rightarrow\infty$. 
${\mathbf X}^N(\cdot/\lambda_N)$ converges weakly to the limit diffusion 
in the space of paths $\omega:[0,\infty )\rightarrow [0,1]^d$, whose 
generator is
\begin{equation}
{\mathcal L}={\mathcal L}_0-b{\mathcal L}_1,
\label{generator}
\end{equation}
where
\begin{eqnarray*}
{\mathcal L}_0&=&\sum_{i=1}^d\frac{x_i(1-x_i)}{2}\frac{\partial^2}{\partial x_i^2}
+\frac{\gamma'}{2}\sum_{i=1}^d(\bar{x}-x_i)\frac{\partial}{\partial x_i}
+\frac{\theta}{2}\sum_{i=1}^d(1-2x_i)\frac{\partial}{\partial x_i},\\
{\mathcal L}_1&=&\frac{\gamma'}{2d}\sum_{i=1}^d\left[(1-2x_i)\sum_{j\neq i}x_j
+(d-1)x_i\right]\frac{\partial}{\partial x_i}.
\end{eqnarray*}
Here, $\bar{x}$ is the arithmetic mean of $\mathbf{x}$ and 
$\gamma'=d\gamma/(d-1)$. 

It is noteworthy that the generator (\ref{generator}) also appears as 
the diffusive limit of the $d$-island model~\cite{Wright1951} with 
allele-dependent migration. The interesting correspondence between a model 
of a multigene family with gene conversion and a $d$-island model was 
previously mentioned in~\cite{ManoInnan2008}. Here, in the current model 
of a multigene family with biased gene conversion, the corresponding 
$d$-island model with allele-dependent migration gives a far more 
intuitive picture of the dual process than the original model of 
a multigene family (see Section 4).

A $d$-island model is a subdivided population consisting of $d$ demes, 
where each deme is occupied by $N$ haploid individuals and 
all pairs of demes can exchange migrants symmetrically. The population 
evolves according to a continuous-time Moran model, in which an individual 
produces one offspring at a time. The allelic type of the offspring is
modified from that of the parent according to the mutation mechanism. 
The offspring then replaces an individual chosen at random from 
the same deme or another deme. The offspring may replace its own parent. 
The replaced individual is removed from the deme, keeping the deme sizes 
constant. We assume that an individual reproduces at a rate 
$\lambda_N$ and replaces an individual of the same deme. In addition, 
during the migration process, alleles $A$ and $a$ replace an individual in 
another deme at rates of $\lambda_N\xi_A$ and $\lambda_N\xi_a$, respectively. 
An offspring will have the same allelic type as the parent with a probability 
of $1-u$ and will have the other type with a probability of $u$. 
$\xi_A=c(1+b)/(d-1)$ and $\xi_a=c(1-b)/(d-1)$ with $0\le b\le 1$, where
$c\,(0<c<1)$ is the migration rate. The migration mechanism has 
allele-dependent bias if $b>0$.

The state of the population at time $t$ can be represented as 
a continuous-time Markov chain $\mathbf{Z}^N(t)=(Z^N_i(t))$, where 
$Z^N_i(t)$ is the number of individuals of allele $A$ in the $i$-th deme at 
time $t$. If $\mathbf{Z}^N(t)=\mathbf{z}$, the transition to 
$\mathbf{z}+\mathbf{e}_i$ occurs at a rate of
\begin{equation}
\lambda_N
(z_i+\xi_A \sum_{k\neq i}z_k)\frac{N-z_i}{N}(1-u)
+\lambda_N
\left[(N-z_i)+\xi_a\sum_{k\neq i}(N-z_k)\right]\frac{N-z_i}{N}u,
\label{trans_island_1}
\end{equation}
and the transition to $\mathbf{z}-\mathbf{e}_i$ occurs at a rate of
\begin{eqnarray}
\lambda_N
\left[(N-z_i)+\xi_a\sum_{k\neq i}(N-z_k)\right]\frac{z_i}{N}(1-u)
+\lambda_N
(z_i+\xi_A\sum_{k\neq i}z_k)\frac{z_i}{N}u.
\label{trans_island_2}
\end{eqnarray}
$X_i(t)=Z_i(t)/N$ denotes the frequency of allele $A$ in the $i$-th deme at 
time $t$. The limiting diffusion approximation is obtained by setting 
$\lambda_N=N/2$ assuming that $Nu\rightarrow\theta$ and 
$Nc\rightarrow\gamma$ as $N\rightarrow\infty$. 
${\mathbf X}^N(\cdot/\lambda_N)$ converges weakly to the limit diffusion in 
$[0,1]^d$, whose generator is identical to the generator of the model of 
biased gene conversion in a multigene family (\ref{generator}).

\section{Fixation probability}

The fate of a single mutant is important in molecular evolutionary
problems. By using a birth-death process, an expression of the fixation
probability of a single mutant in the weak conversion limit 
($\gamma\rightarrow 0$) was obtained~\cite{Walsh1985}. In contrast, when 
$\gamma$ is large, the effects of bias upon the fixation probability can
be significant, as was observed recently obtained using computer 
simulations~\cite{ManoInnan2008}. However, no analytical expression have 
been obtained.

In this section the mutation rate is set to zero (i.e., $u=0$). A path 
of the continuous-time Moran model will eventually be absorbed into 
either of the absorbing states $\mathbf 0$ and $\mathbf 1$, since without 
mutation allele $A$ cannot recover in a population fixed by allele 
$a$, and vice versa. For the diffusive limit of the continuous-time 
Moran model, we have the following lemma.

\begin{lemma}\label{boundary}
The extremal stationary states of the diffusive limit of the continuous-time 
Moran model taking values in $[0,1]^d$ and governed by the generator 
(\ref{generator}) are $\delta_{\mathbf 0}$ and $\delta_{\mathbf 1}$. 
\end{lemma}

\begin{proof}
Let $\nu$ be a extremal stationary state. The system of equations for 
$\mu_{\mathbf{a}}^{\infty}=\langle \nu,{\mathbf X}^{\mathbf a}\rangle$, 
where ${\mathbf X}^{\mathbf a}=\prod_{i=1}^d X_i^{a_i}$ and
$\langle \nu,f\rangle=\int\nu(dx)f(x)$, is obtained by applying It\^o's 
formula to ${\mathbf X}^{\mathbf a}$ with the generator (\ref{generator}).
That is
\begin{eqnarray}
&&\left\{
\sum_{i=1}^d\frac{a_i(a_i-1)}{2}+\frac{\gamma}{2}(1+b)\sum_{i=1}^d a_i
\right\}\mu^{\infty}_{\mathbf a}=
\sum_{i=1}^d\frac{a_i(a_i-1)}{2}\mu^{\infty}_{{\mathbf a}-{\mathbf e}_i}
\nonumber\\
&&+\frac{\gamma'}{d}\sum_{i<j}
a_i\left\{(1-b)\mu^{\infty}_{\mathbf a-{\mathbf e}_i+{\mathbf e}_j}
+2b\mu^{\infty}_{{\mathbf a}+{\mathbf e}_j}\right\}.
\end{eqnarray}
It implies $\mu_{\mathbf a}^{\infty}=\mu_{{\mathbf e}_1}^{\infty}$ for all 
${\mathbf a}\neq {\mathbf 0}$. Thus we have 
$\nu=\mu_{{\mathbf e}_1}^{\infty}\delta_{\mathbf 1}
+(1-\mu_{{\mathbf e}_1}^{\infty})\delta_{\mathbf 0}$.
\end{proof}

\begin{theorem}
For the diffusive limit of the continuous-time Moran model, the fixation 
probability of allele $a$ with ${\mathbf X}(\mathbf 0)=\mathbf{p}$ is
\begin{equation}
\pi(\mathbf{p})=\bar{p}-
\left\{
(d-1)[\bar{p}+\bar{p}(1-\bar{p})\gamma']+\frac{2}{d}
\sum_{i<j}p_ip_j
\right\}b
+O(b^2).
\label{fixprob}
\end{equation}
\end{theorem}

\begin{proof}
Let $\mu_{\mathbf{a}}(t)={\mathbb E}[{\mathbf X(t)}^{\mathbf a}]$.
By Lemma \ref{boundary} and Lebesgue's dominated convergent theorem, 
$\lim_{t\rightarrow\infty}
\mu_{\mathbf{a}}(t)=\mu_{{\mathbf e}_1}^{\infty}=\pi(\mathbf{p})$.
Consider expansion of the Laplace transform of the moments
$\mu_{\mathbf{a}}(t)$ as a power series in $b$:
$\tilde{\mu}_{\mathbf{a}}(s)=\tilde{\mu}_{\mathbf{a}}^{(0)}(s)+\tilde{\mu}_{\mathbf{a}}^{(1)}(s)b
+\cdots$. At the zeroth order in $b$ we have a system of equations
\begin{equation}
(s+\gamma')\tilde{\mu}^{(0)}_{\mathbf{e}_i}
-\frac{\gamma'}{d}\sum_{j=1}^d\tilde{\mu}_{\mathbf{e}_j}^{(0)}=p_i,
\qquad i=1,2,...,n,
\end{equation}
and the solution is
\begin{equation}
\tilde{\mu}^{(0)}_{\mathbf{e}_i}(s)=\frac{\bar{p}}{s}+
\frac{p_i-\bar{p}}{s+\gamma'},\qquad i=1,2,...,d.
\end{equation}
By applying the inverse Laplace transform, we have 
$\mu^{(0)}_{\mathbf{e}_1}=\bar{p}+(p_1-\bar{p})e^{-\gamma' t}$.
In the same manner, for $i=1,2,...,n$,
\begin{equation}
\left(\frac{s}{2}+1+\gamma'\right)\tilde{\mu}^{(0)}_{2\mathbf{e}_i}
-\tilde{\mu}^{(0)}_{\mathbf{e}_i}
-\frac{\gamma'}{d}\sum_{j=1}^n\tilde{\mu}^{(0)}_{\mathbf{e}_i+\mathbf{e}_j}
=\frac{p^2_i}{2}
\end{equation}
and for $i\neq j;~i,j=1,2,...,n$,
\begin{equation}
(s+2\gamma')\tilde{\mu}^{(0)}_{\mathbf{e}_i+\mathbf{e}_j}
-\frac{\gamma'}{d}\sum_{k=1}^n
(\tilde{\mu}^{(0)}_{\mathbf{e}_j+\mathbf{e}_k}+\tilde{\mu}^{(0)}_{\mathbf{e}_i+\mathbf{e}_k})
=p_ip_j.
\end{equation}
They can be solved for $\tilde{\mu}^{(0)}_{2\mathbf{e}_i}$ and 
$\tilde{\mu}^{(0)}_{\mathbf{e}_i+\mathbf{e}_j}$. At the first order in $b$, we have 
a system of equations for $i=1,2,...,d$,
\begin{equation}
(s+\gamma')\tilde{\mu}^{(1)}_{\mathbf{e}_i}
-\frac{\gamma'}{d}\sum_{j=1}^n\tilde{\mu}_{\mathbf{e}_j}^{(1)}
=\frac{\gamma'}{d}\left\{(d-2)\tilde{\mu}^{(0)}_{\mathbf{e}_i}
+\sum_{j=1}^d\tilde{\mu}^{(0)}_{\mathbf{e}_j}
-2\sum_{j(\neq i)}\tilde{\mu}_{\mathbf{e}_i+\mathbf{e}_j}^{(0)}\right\}.
\end{equation}
Then,
\begin{equation}
\tilde{\mu}^{(1)}_{\mathbf{e}_i}(s)=
\frac{a_0}{s}+\sum_{j=1}^d\frac{a_j}{s-s_j},\qquad i=1,2,...,d,
\end{equation}
where
\begin{equation}
a_0=(d-1)\bar{p}\{1+\gamma'(1-\bar{p})\}
-\frac{2}{d}\sum_{i<j}p_ip_j.
\end{equation}
$s_j~(<0)$ are eigenvalues of the generator (\ref{generator}) and 
$a_{j\neq 0}$ are constants independent of $s$. Then, by applying the inverse 
Laplace transform, the lemma follows.
\end{proof}

\begin{remark}
In the weak conversion limit ($\gamma\rightarrow 0$), the expression for
the fixation probability (\ref{fixprob}) with ${\mathbf p}={\mathbf e}_1/N$ 
in large $N$ agrees with Equation 8 of~\cite{Walsh1985} which was obtained
by a different method. 
\end{remark}

\begin{remark}\label{remark_fix_weakconv}
It may seems curious that the effects of bias (linear term in $b$) do not 
vanish in the weak conversion limit ($\gamma\rightarrow 0$). Of course, 
the linear term disappears without gene conversion ($c=0$). If gene 
conversion is extremely weak, all loci are monomorphic except for very 
short periods of time when a single locus is segregating. An allele fixes 
in the polymorphic locus and after a long period of time biased gene 
conversion creates another polymorphic locus. The process continues until 
all loci are fixed by the same allele. Since the locus-by-locus spreading is
biased, the bias is effective even when gene conversion is extremely weak
(see~\cite{Walsh1985}).
\end{remark}

When $b=1$ all conversion events involving different alleles result in
an allele $a$ being converted by an allele $A$.

\begin{theorem}
When $b=1$, 
\begin{equation}
\pi(\mathbf p)=\left\{1-(1-\bar{p})d\gamma+O(\gamma^2)\right\}\prod_{i=1}^dp_i.
\end{equation}
\end{theorem}

\begin{proof}
The fixation probability satisfies the Kolmogorov backward equation
\begin{equation}
({\mathcal L}_0-{\mathcal L}_1)\pi({\mathbf x})=0
\label{kbe_b=1}
\end{equation}
with $\pi({\mathbf 0})=0$ and $\pi({\mathbf 1})=1$. Assuming
$\pi({\mathbf x})=\sum_{\mathbf a}c_{\mathbf a}{\mathbf x}^{\mathbf a}$,
(\ref{kbe_b=1}) gives $c_{\mathbf 1}=1-d\gamma+O(\gamma^2)$,
$c_{{\mathbf 1}+{\mathbf e}_i}=\gamma c_{\mathbf 1}$, and
$c_{\mathbf a}=O(\gamma^2)$ for
${\mathbf a}\neq{\mathbf 1},{\mathbf 1}+{\mathbf e}_i$,
where $i=1,2,...,d$.
\end{proof}

\section{Strong conversion limit}

Let us define strong conversion limit with allele dependent migration as 
$dNb\rightarrow\beta$ as $N\rightarrow\infty$. Interestingly, the strong
conversion limit of the continuous-time Moran model of biased gene 
conversion within a $d$-unlinked multigene family has the limiting 
diffusion whose generator is identical to that of the very fundamental 
one-locus diffusion with directional selection. Various results known for 
the process also hold for the strong conversion limit of the biased gene 
conversion model. 

By applying the singular perturbation theory, \cite{EthierNagylaki1980} 
obtained a diffusion approximation of Markov chains with two time scales. 
Consider the continuous-time Moran model for the $d$-island population 
subdivision with allele-dependent migration, whose transition rates are 
(\ref{trans_island_1}) and (\ref{trans_island_2}). Consider the mean of 
the frequencies $\bar{X}^N(t)=\sum_{i=1}^d X^N_i(t)/d$ and the deviations 
from the mean $Y_i^N(t)=X^N_i(t)-\bar{X}^N(t),\, i=1,2,...,d$.
Set $\epsilon_N^{-1}=dN/2/(1+c)$ and $\delta_N^{-1}=1$. Asymptotically 
$N\rightarrow\infty$, the infinitesimal variances and means of
$\bar{X}^N(\cdot/\epsilon_N)$ are
\begin{eqnarray}
&&\epsilon_N^{-1}{\mathbb E}[\bar{X}^N(1)-\bar{x}]=
m(\bar{x},{\mathbf y})+o(1),\\
&&\epsilon_N^{-1}{\mathbb E}[(\bar{X}^N(1)-\bar{x})^2]=
v(\bar{x},{\mathbf y})+o(1),
\end{eqnarray}
where
\begin{eqnarray}
&&m(\bar{x},{\mathbf y})=\frac{\beta c}{1+c}
\left\{\bar{x}-\frac{d}{d-1}
\left[\bar{x}^2-\frac{1}{d^2}\sum_{i=1}^d(y_i+\bar{x})^2\right]\right\}
+\frac{d\theta}{2}(1-2\bar{x}),\\
&&v(\bar{x},{\mathbf y})=\bar{x}-\frac{c}{1+c}\frac{d}{d-1}\bar{x}^2
+\frac{c-d+1}{(1+c)d(d-1)}\sum_{i=1}^d(y_i+\bar{x})^2.
\end{eqnarray}
Those of ${\mathbf Y}^N(\cdot/\delta_N)$ are 
\begin{eqnarray}
&&\delta_N^{-1}{\mathbb E}[{\mathbf Y}^N(1)-{\mathbf y}]
=f(\bar{x},{\mathbf y})+o(1),
\end{eqnarray}
where $f(\bar{x},{\mathbf y})=-nc\mathbf{y}/(d-1)$. Also,
$\epsilon_N^{-1}{\mathbb E}[(\bar{X}^N(1)-\bar{x})^4]=o(1)$,
$\delta_N^{-1}{\rm Var}[{\mathbf y}^N(1)]=o(1)$.
The zero solution of
\begin{equation} 
\frac{d\mathbf{y}}{dt}=f(\bar{x},{\mathbf y})
\end{equation}
is globally asymptotically stable. Then, according to Theorem 3.3 
of~\cite{EthierNagylaki1980}
${\mathbf Y}^N(t/\epsilon_N)\rightarrow \mathbf 0$ in probability for
every $t>0$, and $\bar{X}^N(\cdot/\epsilon_N)$ converges weakly to 
a diffusion process on the surface ${\mathbf y}={\mathbf 0}$. 

\begin{theorem}\label{stlimit}
The continuous-time Moran model for the $d$-island population subdivision 
with allele-dependent migration whose transition rates are
(\ref{trans_island_1}) and (\ref{trans_island_2}) has the limiting 
diffusion of strong migration in $[0,1]$ with a generator
\begin{equation}
\frac{{\bar x}(1-{\bar x})}{2}\frac{\partial^2}{\partial {\bar x}^2}
-\left[\frac{\beta c}{1+c}{\bar x}(1-{\bar x})-\frac{d\theta}{2}(1-2{\bar x})
\right]\frac{\partial}{\partial {\bar x}}.
\label{stgenerator}
\end{equation}
This generator also appears as the strong conversion limit of 
the continuous-time Moran model of biased gene conversion within 
a $d$-unlinked multigene family whose transition rates are 
(\ref{trans_multi_1}) and (\ref{trans_multi_2}).
\end{theorem}

\begin{remark}
The generator (\ref{stgenerator}) is identical to that of the diffusion 
process of the one-locus two-allele model with directional selection with
selection intensity $2\beta c/(1+c)$. In strong migration limits of 
population subdivision considered by~\cite{Nagylaki1980}, the effects of 
population subdivision disappear and the panmictic diffusion holds if 
the migration is conservative. In contrast, in the strong conversion limit of 
the continuous-time Moran model of biased gene conversion, the effects of 
a multigene structure remain as an effective selection. 
\end{remark}

\begin{remark}\label{convASG}
The continuous-time Moran model for one-locus two-alleles with directional 
selection can be formulated by the biased voter model and the dual has 
a limit process that generates the ancestral selection 
graph~\cite{KroneNeuhauser1997}. The continuous-time Moran model of biased 
gene conversion has an analogue of the ancestral selection graph, which
we call the ancestral bias graph (see Section 5). In the strong 
conversion limit, the process generating the ancestral bias graph should 
converge into the process generating ancestral selection graph with
selection intensity $2\beta c/(1+c)$. A direct proof of this observation
without using the duality argument seems difficult.
\end{remark}

\section{Ancestral bias graph}

The above introduced $d$-island model with allele-dependent migration also 
has a formulation in terms of the biased voter model on a set of complete 
graphs. Let $\mathbf{I}=(I_i),~I_i=\{1,2,...,N\},~i=1,2,...,d$ denotes sets 
of sites, where $I_i$ is the set of sites in the $i$-th graph. The biased 
voter model is a continuous-time Markov process whose state at time $t$ is 
denoted by $\eta_t:~\mathbf{I}\rightarrow\{A,a\}$. If $x\in I_i,~\eta_t(x)=A$
($a$), then $x$ is occupied by an individual of allelic type
$A$ ($a$) at time $t$. The process $\{\eta_t;~t\ge 0\}$ evolves 
according to the following rules.

\begin{enumerate}

\item For $x=1,2,...,N$ and $i=1,2,...,d$, the individual at $x\in I_i$ 
produces an offspring at rate of $\lambda_N$ within $I_i$.

\item The offspring has the same allelic type as the parent with a probability
of $1-u$ and has the other type with a probability of $u$.

\item For $x=1,2,...,N,~j\neq i;~i,j=1,2,...,d$, the individual at 
$x \in I_i$ produces an offspring in $I_j$ at rates depending on the allelic
type. If $\eta_t(x)=A$ ($a$), the rate is $\lambda_N\xi_A$ ($\lambda_N\xi_a$).
$\xi_A-\xi_a=2cb/(d-1)$. 

\item At the time when the birth event occurs, one of the $N$ sites is chosen
at random and the individual at this site is replaced by the offspring.
The offspring is allowed to replace its own parent. 

\end{enumerate}

This process can be visualized by a percolation process 
\cite{Harris1972,Donnelly1984}, and the construction is similar to that of 
the continuous-time Moran model with selection~\cite{KroneNeuhauser1997}. 
The idea is to construct the process using a collection of independent 
Poisson processes by drawing arrows on the space-time coordinate system 
$\mathbf{I}\times[0,\infty)$. These arrows indicate where and when 
the offspring is produced and sent. We begin by connecting 
arrows to each time-line at the times of arrivals in a Poisson process that 
describes the birth process. For each $(x,y)\in I_i^2,~i=1,2,...,d$, let 
$\{W_{i,s}^{x,y};~s\ge 1\}$ denote the times of arrivals in a Poisson 
process with rate $\lambda_N/N$. For each 
$(x,y)\in I_i\times I_j, i\neq j, i,j=1,2,...,d$, let 
$\{Z_{i,j,s}^{x,y};~s\ge 1\}$ denote times of arrivals in a Poisson process
with rate $\lambda_N\xi_A/N$. Let $\{U_{i,j,s}^{x,y};~s\ge 1\}$ and 
$\{V_{i,s}^{x,y};~s\ge 1\},~i\neq j,~i,j=1,2,...,d$ be sequences of 
independent, uniformly distributed random variables in $(0,1)$. For times 
$W_{i,s}^{x,y}$ we draw an arrow from $x\in I_i$ to $y\in I_i$ to indicate 
the birth of an offspring at $x$ that is sent to $y$. For times 
$Z_{i,j,s}^{x,y}$ we draw an arrow from $x\in I_i$ to $y\in I_j$ to indicate
the birth of an offspring at $x$ that is then sent to $y$. 
If $U_{i,j,s}^{x,y}<\xi_a/\xi_A$, we place a ``$\delta$'' at the tip of 
the arrow; otherwise, we label the arrow with a ``2''. In other words, we have 
$\delta$-arrows and 2-allows entering a site $y$ at rates $\lambda_N\xi_a$ and
$\lambda_N(\xi_A-\xi_a)$, respectively. Then, the following rule will 
apply: type $A$ individuals can give birth through both types of arrows, but 
type $a$ individuals can only give birth through $\delta$-arrows. The process
$\{V_{i,s}^{x,y};~s\ge 1\}$ is used as the mutation process; if 
$V_{i,s}^{x,y}<u$, a mutation occurs. We represent a mutation event by solid 
dots on the arrows. A realization of the percolation diagram in the case 
$d=2$ and $N=4$ is shown in Fig. 1. $I_{1}$ and $I_{2}$ are the left and 
the right graphs, respectively. If the set of sites occupied by type $a$ 
individuals initially is $\{1\}\in I_{2}$, then, at time $t$, the set of 
sites occupied by type $a$ individuals is $\{2\}\in I_{1}$ and 
$\{3\}\in I_{2}$. The paths of the $a$'s are indicated by thick lines.
By reversing time, the ancestral history of individuals at sites are followed
and thus their types are determined. The resulting process is called the dual
process. A realization of the dual process, which was obtained from Fig. 1 by
simply reversing time and the direction of the arrows, is shown in Fig. 2. 
Here, the ancestral history of a sample consists of individuals at sites 
$\{1,2\}\in I_{1}$, and $\{1\}\in I_{2}$ at dual time 0 is indicated by thick
lines.

Consider the dual process of a sample of size 
${\mathbf n}=(n_i),\,i=1,2,...,d$ at time $0$ taken from a population, when 
the mutation rate is set to zero (i.e., $u=0$). Assume there are ${\mathbf k}$
particles in the limiting process. A {\em coalescing event} occurs when 
a particle crosses an unmarked arrow and lands on the site of a different 
particle contained in the dual process. This occurs at rates
\begin{equation}
\lambda_Nk_i\frac{k_i-1}{N}=\frac{k_i(k_i-1)}{2},\qquad i=1,2,...,d.
\label{abg1}
\end{equation}
A {\em migration event} occurs when a particle crosses a $\delta$-arrow. 
This occurs at a rate of 
\begin{equation}
\lambda_N\xi_an_i\frac{N-k_j}{N}\rightarrow\frac{(1-b)\gamma}{2(d-1)}k_i,
\qquad N\rightarrow\infty,
\label{abg2}
\end{equation}
for $j\neq i; i,j=1,2,...,d$. A {\em branching event} occurs when
a particle crosses a 2-arrow. This occurs at a rare of
\begin{equation}
\lambda_N(\xi_A-\xi_a)k_i\frac{N-k_j}{N}\rightarrow\frac{b\gamma}{d-1}k_i,
\qquad N\rightarrow\infty,
\label{abg3}
\end{equation}
for $i\neq j; i,j=1,2,...,d$. The original particle continues along the
continuing path, and the new particle that arose from the branching follows
the 2-allow (incoming path). The new particle can land on a site that is
already contained in the dual process.  The event, which was called 
{\em collision} by \cite{KroneNeuhauser1997}, occurs with a probability 
of $k_j/N$, and can be ignored in the limit $N\rightarrow\infty$. An analogue
of the coalescent genealogy can be obtained by rescaling time and 
the parameters as $\lambda_N=N/2, \xi_A=c(1+b)/(d-1), \xi_a=c(1-b)/(d-1)$ with 
$Nc\rightarrow\gamma$ as $N\rightarrow\infty$. We call the limiting process 
$\{{\mathcal G}_{\mathbf n}(t);t\ge 0\}$ the {\em ancestral bias graph}, which
consists of three components: the set valued process 
$\{{\mathcal A}_{{\mathbf n},i}(t);t\ge 0\;i=1,2,...,d\}$, the jump process 
$\{R_m; m\ge 1\}$, and the label process $\{(\beta_m,\gamma_m); m\ge 1\}$.
${\mathcal A}_{{\mathbf n},i}(t)$ is the set of particles in $i$-th deme at time
$t$, and $R_m$ is the time of the $m$-th event. $\beta_m$ and $\gamma_m$
denotes branched particles or coalesced particles at the $m$-th event, 
respectively. 

Let the size process 
$\mathbf{A}_{\mathbf n}(t)=(|{\mathcal A}_{{\mathbf n},i}(t)|),~i=1,2,...,d$.
$\{\mathbf{A}_{\mathbf n}(t);t\ge 0\}$ is a $d$-dimensional birth and death 
process with rates (\ref{abg1}--\ref{abg3}) in the state space 
${\mathbb Z}_+^d$ . The process is similar to the process that was introduced
by \cite{ShigaUchiyama1986} to analyze stationary states and their 
stability of the stepping stone model involving directional selection.
By either a direct application of It\^o's formula or using rules to generate 
an ancestral bias graph for a sample that consists of allele $a$, we obtain
the following duality relation. The proofs are essentially the same as 
that of Theorem 2.1 of \cite{Mano2009} for the size process of the ancestral
selection graph.

\begin{lemma}\label{duality}
The moment dual of the birth and death process $\mathbf{A}_{\mathbf n}(t)$
is the Wright-Fisher diffusion $\mathbf{X}(t)$ in $[0,1]^d$ governed by 
the generator (\ref{generator}).
\begin{equation}
{\mathbb E}_{\mathbf{p}}\left[{{\mathbf X}(t)}^{\mathbf n}\right]
={\mathbb E}_{\mathbf{n}}\left[{\mathbf p}^{{\mathbf A}_{\mathbf n}(t)}
\right].
\label{duality_eq}
\end{equation}
\end{lemma}

\begin{corollary}
The joint probability generating function of the stationary measure of 
the birth and death process ($\pi_A$) with rates 
{\rm (\ref{abg1}--\ref{abg3})} is
\begin{equation}
{\mathbb E}_{\pi_A}\left[{\mathbf p}^{\mathbf A}\right]=\pi(\mathbf{p}),
\end{equation}
where $\pi(\mathbf{p})$ is the fixation probability given in Theorem 3.1.
In particular,
\begin{eqnarray}
&&\pi_A(\mathbf{e}_i)=\frac{1}{d}-b\left(\frac{d-1}{d}-\gamma\right)+O(b^2),
\label{stmeasure1}\\
&&\pi_A
(2\mathbf{e}_i)=b\frac{\gamma}{d}+O(b^2),\\
&&\pi_A
(\mathbf{e}_i+\mathbf{e}_j)=b\frac{2}{d}(1+\gamma)+O(b^2),
\label{stmeasure3}
\end{eqnarray}
for $i\neq j;~i,j=1,2,...,d$. Other configurations have probabilities of 
$O(b^2)$. 
\end{corollary}

If the process hits $|{\mathbf A}_{\mathbf n}(t)|=1$ for the first time,
we call the particle at that time the {\em ultimate ancestor}. When $0<b<1$
the ultimate ancestor always exists because the state space is irreducible. 
In contrast, when $b=1$, the ultimate 
ancestor never exist as long as $\#\{i;n_i\ge 1\}\ge2$. The state space is 
reducible and states $n_i=0$ for some $i$ are transient. Therefore, if 
$n_i=0$ for some $i$, $\pi_A({\mathbf n})=0$. Moreover, the fixation 
probability given in 
Theorem 3.2 gives

\begin{corollary}
When b=1,
\begin{eqnarray}
&&\pi_A({\mathbf 1})=1-d\gamma+O(\gamma^2),\\
&&\pi_A({\mathbf 1}+{\mathbf e}_i)=\gamma\pi_A({\mathbf 1}).
\end{eqnarray}
for $i=1,2,...,d$. Other configurations have probabilities $O(\gamma^2)$.
\end{corollary}

\begin{theorem}
Let $W_{\mathbf n}$ be the waiting time to the ultimate ancestor of a sample
of ${\mathbf n}$ genes. 
${\mathbb E}[W_{{\mathbf e}_1}]=0$ and
\begin{eqnarray}
r(\mathbf n){\mathbb E}[W_{\mathbf n}]&=&
\sum_{i=1}^d n_i(n_i-1){\mathbb E}[W_{{\mathbf n}-{\mathbf e}_i}]
+\sum_{i\neq j}\frac{2b\gamma n_i}{d-1}
{\mathbb E}[W_{{\mathbf n}+{\mathbf e}_j}]
\nonumber\\
&&+\sum_{i\neq j}\frac{(1-b)\gamma n_i}{d-1}
{\mathbb E}[W_{{\mathbf n}-{\mathbf e}_i+{\mathbf e}_j}]+2,
\end{eqnarray}
where $r(\mathbf{n})=\sum_{i=1}^d n_i\{n_i-1+\gamma(1+b)\}$.
\end{theorem}

\begin{proof}
This is clear by considering the waiting time until the first event in 
the birth and death process with rates (\ref{abg1}-\ref{abg3}).
\end{proof}

\begin{corollary}
For a sample of size two,
\begin{eqnarray}
&&{\mathbb E}[W_{2{\mathbf e}_i}]
=d\nonumber\\
&&
+b\frac{2(d\gamma)^3+
(d-1)\{8(d\gamma)^2+(10d^2+7d-2)\gamma+2(d-1)(2d-1)\}}
{3\gamma\{d\gamma+2(d-1)\}}+O(b^2),\nonumber\\
\\
&&{\mathbb E}[W_{{\mathbf e}_i+{\mathbf e}_j}]
=d+\frac{d-1}{\gamma}+b\frac{2d^2\gamma^3+2(d-1)\{3d^2\gamma+2(d^2-1)\}}
{3\gamma\{d\gamma+2(d-1)\}}+O(b^2),
\end{eqnarray}
for $i\neq j;i,j=1,2,...,d$.
\end{corollary}

Consider superimposing the mutation process, which is denoted by 
$\{{\mathcal Y}_i;t\ge 0,i=1,2,...,d\}$, on the ancestral bias graph. The rate is 
$\lambda_N uk_i\rightarrow\theta k_i/2,\, N\rightarrow\infty$, $i=1,2,...,n$.
Depending on the type of the ultimate ancestor and the mutation events along 
the branches, certain parts of the ancestral bias graph may not be accessible 
to individuals since only individuals of allelic type $A$ may cross 2-arrows. 
In Fig. 2, if the allelic type of the ultimate ancestor is $a$, then the true 
genealogy contains the dotted line and does not contain the dashed line. 
In contrast,if the the allelic type of the ultimate ancestor is $A$, then  
the true genealogy contains the dashed line. To simulate the joint 
distribution of a sample of size $\mathbf n$ from a large population that is
in equilibrium, we proceed as follows:

\begin{enumerate}

\item{Construct an ancestral bias graph starting with $\mathbf n$ particles, 
until the first time it reaches the ultimate ancestor.}

\item{Choose the type of the ultimate ancestor of the sample according to 
the stationary measure of the diffusion process governed by the generator 
(\ref{generator}) (see Section 6).}

\item{Run the mutation process forward along the ancestral bias graph starting 
at the ultimate ancestor.}

\end{enumerate}

In step (iii), the type of particle continues after the meeting of incoming 
and continuing branches (branching event in the dual process) and is identical
to Table 2 of \cite{KroneNeuhauser1997}, in which types `1' and `2' are `a' 
and `A', respectivel.

As for the ancestral selection graph~\cite{KroneNeuhauser1997}, the
effects of bias on the ancestral bias graph when imposing the mutation process
are insignificant when the mutation rate is very large or very small. 

\begin{lemma}\label{lemma_TMRCA}
Let time to the most recent common ancestor be $T_{MRCA}$. When $\theta=0$, 
the distribution of $T_{MRCA}$ does not depend on $b$ and is identical to 
that of the $d$-island model of population subdivision without bias. When 
$\theta\gg b\gamma$, the distribution of $T_{MRCA}$ is also identical to 
the model without bias.
\end{lemma}

\begin{proof}
The proof is essentially the same as that of Theorem 3.12 of 
\cite{KroneNeuhauser1997}.
\end{proof}

\section{Sampling distributions}

Suppose that $\mathbf{n}$ genes are sampled from a population. A sample 
configuration is denoted by $(\mathbf{n}_a,\mathbf{n}_A)$, where 
$\mathbf{n}=\mathbf{n}_a+\mathbf{n}_A$, and $n_{a,i}$ and $n_{A,i}$ are 
the number of allele $a$ and of allele $A$ in the $i$-th deme, respectively. 
Let $p(\mathbf{n})$ be the the multinomial sampling distribution, 
or the likelihood, of a sample of $\mathbf{n}$ genes taken from a population 
in the equilibrium,
\begin{equation}
p(\mathbf{n})=
{\mathbb E}_{\pi_{\mathbf X}}
\left[\prod_{i=1}^d\frac{n_i!}{n_{a,i}!n_{A,i}!}
X_i^{n_{a,i}}(1-X_i)^{n_{A,i}}\right]. 
\label{def_like}
\end{equation}

\begin{theorem}\label{sampdist}
The sampling distribution $p(\mathbf{n})$ satisfies 
\begin{eqnarray}
&&r(\mathbf{n})p(\mathbf{n})
=\sum_{i=1}^d
\left[(n_{a,i}-1)n_ip(\mathbf{n}-\mathbf{e}_{a,i})
+(n_{A,i}-1)n_ip(\mathbf{n}-\mathbf{e}_{A,i})\right]
\nonumber\\
&&+\theta\sum_{i=1}^d
\left[(n_{A,i}+1)p(\mathbf{n}_a-\mathbf{e}_{a,i}+\mathbf{e}_{A,i})
+(n_{a,i}+1)p(\mathbf{n}+\mathbf{e}_{a,i}-\mathbf{e}_{A,i})\right]
\nonumber\\
&&+\frac{(1-b)\gamma'}{d}\sum_{i,j\neq i}n_i
\left[
\frac{n_{a,j}+1}{n_j+1}p(\mathbf{n}-\mathbf{e}_{a,i}+\mathbf{e}_{a,j})
+\frac{n_{A,j}+1}{n_j+1}p(\mathbf{n}-\mathbf{e}_{A,i}+\mathbf{e}_{A,j})
\right]\nonumber\\
&&+\frac{2b\gamma'}{d}\sum_{i,j\neq i}
\left[
\frac{n_{A,i}(n_{A,j}+1)}{n_j+1}p(\mathbf{n}+\mathbf{e}_{A,j})
+\frac{n_i(n_{a,j}+1)}{n_j+1}p(\mathbf{n}+\mathbf{e}_{a,j})
\right.
\nonumber\\
&&\left.
+\frac{(n_{a,i}+1)(n_{A,j}+1)}{n_j+1}
p(\mathbf{n}+\mathbf{e}_{a,i}-\mathbf{e}_{A,i}+\mathbf{e}_{A,j})\right],
\label{rec}
\end{eqnarray}
where $r(\mathbf{n})=\sum_{i=1}^d n_i\{n_i-1+\theta+\gamma(1+b)\}$.
The probabilities with negative arguments are zero. The boundary condition is
\begin{equation}
p(\mathbf{e}_{a,i})=\rho,\qquad p(\mathbf{e}_{A,i})=1-\rho,
\qquad i=1,2,...,d.
\label{bcr}
\end{equation}
\end{theorem}

\begin{proof}
It follows by a direct application of It\^o's formula to the moments in 
(\ref{def_like}). It can also be proved using the rules to generate 
the ancestral bias graph when imposing the mutation 
process~\cite{KroneNeuhauser1997}.
\end{proof}

\begin{remark}
In the weak mutation limit ($\theta\rightarrow 0$), 
$p(\mathbf{n}_a)=p(\mathbf{e}_1)=\rho$ for any nonzero
$\mathbf{n}_a\in
{\rm span}\{\mathbf{e}_{a,1},\mathbf{e}_{a,2},...,\mathbf{e}_{a,d}\}$,
$p(\mathbf{0},\mathbf{n}_A)=1-\rho$ for any nonzero 
$\mathbf{n}_A\in
{\rm span}\{\mathbf{e}_{A,1},\mathbf{e}_{A,2},...,\mathbf{e}_{A,d}\}$, 
and $p(\mathbf{n})=0$ for any nonzero
$\mathbf{n}\notin
\{{\rm span}\{\mathbf{e}_{a,1},\mathbf{e}_{a,2},...,\mathbf{e}_{a,d}\}$,
${\rm span}\{\mathbf{e}_{A,1},\mathbf{e}_{A,2},...,\mathbf{e}_{A,d}\}\}$,
because the sample allelic type is solely determined by the allelic 
type of the ultimate ancestor. For example, if a sample of size two 
is taken from the $i$-th deme, 
$p(2\mathbf{e}_{a,i})+p(2\mathbf{e}_{A,i})=1$ and 
$p(\mathbf{e}_{a,i}+\mathbf{e}_{A,i})=0$.
\end{remark}

\begin{corollary}
For a sample of size one,
\begin{equation}
p(\mathbf{e}_{a,i})=\frac{1}{2}-
\frac{(1+2\theta+\gamma')\gamma}
{2\{\gamma'/d+2\theta(1+2\theta+\gamma')\}}b+O(b^2),\qquad
i=1,2,...,d,
\end{equation}
and for a sample of two,
\begin{eqnarray}
&&p(2\mathbf{e}_{a,i})=
\frac{\gamma'/d+\theta+(1+2\theta+\gamma')(\theta-b\gamma)}
{2\{\gamma'/d+2\theta(1+2\theta+\gamma')\}}+O(b^2),
\label{mom1}\\
&&p(\mathbf{e}_{a,i}+\mathbf{e}_{a,j})=
\frac{\gamma'/d+(1+2\theta+\gamma')(\theta-b\gamma)}
{2\{\gamma'/d+2\theta(1+2\theta+\gamma')\}}+O(b^2),
\label{mom2}\\
&&p(\mathbf{e}_{a,i}+\mathbf{e}_{A,i})=
\frac{\theta(2\theta+\gamma')}{\gamma'/d+2\theta(1+2\theta+\gamma')}+O(b^2),
\\
&&p(\mathbf{e}_{a,i}+\mathbf{e}_{A,j})=
\frac{\theta(1+2\theta+\gamma')}
{2\{\gamma'/d+2\theta(1+2\theta+\gamma')\}}+O(b^2).
\end{eqnarray}
for $i\neq j;~i,j=1,2,...,d$. 
$p(2\mathbf{e}_{A,i})$ and $p(\mathbf{e}_{A,i}+\mathbf{e}_{A,j})$ 
are given by (\ref{mom1}) and (\ref{mom2}), respectively, by replacing $b$ 
with $(-b)$. These expressions reduce to those in the model without bias 
when $\theta\gg b\gamma$ (see Lemma \ref{lemma_TMRCA}). The effects of 
bias (linear term in $b$) vanish in the weak conversion limit 
($\gamma\rightarrow 0$) (c.f. Remark \ref{remark_fix_weakconv}), but
the effects do not vanish in the weak mutation limit 
(c.f. Lemma \ref{lemma_TMRCA}).
\end{corollary}

\begin{remark}
The identity coefficients in a multigene family were defined 
by~\cite{Ohta1982}. For unlinked loci, the average probability of 
identity at the same locus is
$p(2\mathbf{e}_{a,i})+p(2\mathbf{e}_{A,i})$,
and that at different loci of the same or homologous chromosomes is
$p(\mathbf{e}_{a,i}+\mathbf{e}_{a,j})+p(\mathbf{e}_{A,i}+\mathbf{e}_{A,j})$.
When $b=0$, these expressions reduce to Equations~12 of~\cite{Ohta1982}.
\end{remark}

\section{Importance sampling}

The state space of ancestral histories of a sample is huge and closed form
expressions for the likelihood are not available. \cite{GriffithsTavare1994}
introduced an importance-sampling method on the ancestral process
back in time. \cite{StephensDonnelly2000} constructed an efficient proposal 
distribution, and \cite{DeIorioGriffiths2004a} characterized the proposal 
distribution in terms of the generator of the dual diffusion process, which
describes the population gene frequencies. \cite{DeIorioGriffiths2004b}
applied the method to construct an importance-sampling algorithm for 
computing the likelihood of samples in subdivided population models.

A history $\{H_k;~k=0,-1,...,-m\}$ is defined as the set of ancestral 
configurations at the embedded events in the Markov process where 
coalescence, migration, branching, and mutation events take place. 
$H_0$ denotes the current state (sample configuration), and $H_{-m}$ 
the state when an ultimate ancestor is reached ($\mathbf{e}_{A,i}$
or $\mathbf{e}_{a,i}$). The system of equations of Theorem~\ref{sampdist} is 
written as $p(H_k)=\sum_{\{H_{k-1}\}}p(H_k|H_{k-1})p(H_{k-1})$.
The importance-sampling representation is based on finding a good 
approximation to the reverse chain probabilities $\hat{p}(H_{k-1}|H_k)$.
The importance-sampling representation is then
\begin{equation}
p(H_0)={\mathbb E}_{\hat{p}}
\left[
\frac{p(H_0|H_{-1})}{\hat{p}(H_{-1}|H_0)}\cdots
\frac{p(H_{-m+1}|H_{-m})}{\hat{p}(H_{-m}|H_{-m+1})}p(H_{-m})
\right]
\approx\frac{1}{M}\sum_{i=1}^M
\frac{p({\mathcal H}^{(i)})}{\hat{p}({\mathcal H}^{(i)})}
\hat{p}(H_0)
\label{IS_sim}
\end{equation}
where ${\mathbb E}_{\hat{p}}$ denotes expectation taken over histories in
the reverse direction with the reverse chain transition probabilities 
$\hat{p}(H_{k-1}|H_k)$, and 
${\mathcal H}^{(1)},{\mathcal H}^{(2)},...,{\mathcal H}^{(M)}$ are independent 
sample paths from the reverse chain.

From (\ref{def_like}) it follows that
\begin{eqnarray}
&&\pi(\alpha|i,{\mathbf n})p(\mathbf n)
={\mathbb E}_{\pi_{\mathbf X}}
\left[(X_i\delta_{\alpha,a}+(1-X_i)\delta_{\alpha,A})
q_\mathbf{n}(\mathbf X)\right],
\label{IS_pi1}\\
&&\pi(\beta|j,{\mathbf n}+{\mathbf e}_{a,i}\delta_{\alpha,a}+
{\mathbf e}_{A,i}\delta_{\alpha,A})
\pi(\alpha|i,{\mathbf n})p(\mathbf n)\nonumber\\
&&={\mathbb E}_{\pi_{\mathbf X}}
\left[(X_j\delta_{\beta,a}+(1-X_j)\delta_{\beta,A})
(X_i\delta_{\alpha,a}+(1-X_i)\delta_{\alpha,A})
q_\mathbf{n}(\mathbf X)\right],
\label{IS_pi2}
\end{eqnarray}
where 
\begin{equation}
q_{\mathbf{n}}(\mathbf x)=\prod_{i=1}^d\frac{n_i!}{n_{a,i}!n_{A,i}!}
x_i^{n_{a,i}}(1-x_i)^{n_{A,i}}
\end{equation}
and $\pi(\alpha|i,{\mathbf n})$ is the probability that an additional 
gene taken from deme $i$ is of allelic type $\alpha$ given that 
we have a configuration $\mathbf n$. Assume (\ref{IS_pi1}) and 
(\ref{IS_pi2}) hold for an approximate sampling distribution 
$\hat{p}(\mathbf n)$ obtained by setting~\cite{DeIorioGriffiths2004b}
\begin{equation}
{\mathbb E}_{\pi_{\mathbf X}}
\left[{\mathcal L}_i\frac{\partial}{\partial X_i} q_{\mathbf n}(\mathbf X)
\right]=0,
\end{equation}
where
\begin{equation}
{\mathcal L}_i=\frac{x_i(1-x_i)}{2}\frac{\partial}{\partial x_i}
+\frac{\gamma'}{2}(\bar{x}-x_i)
+\frac{\theta}{2}(1-2x_i)-b\frac{\gamma'}{2d}
\left[(1-2x_i)\sum_{j\neq i}x_j+(d-1)x_i\right],
\end{equation}
yields a system of equations for $\hat{\pi}(\alpha|i,{\mathbf n})$, which is
an approximation of $\pi(\alpha|i,{\mathbf n})$ for $\alpha=a,A;~i=1,2,...,d$:
\begin{equation}
r_i\hat{\pi}(a|i,{\mathbf n})=n_{a,i}+\theta
+\frac{(1-b)\gamma'}{d}\sum_{j\neq i}\hat{\pi}(a|j,\mathbf{n})
+\frac{2b\gamma'}{d}\frac{n_{A,i}(n_i+1)}{n_{a,i}+1}\hat{\pi}(a|i,\mathbf{n})
\sum_{j\neq i}\hat{\pi}(A|j,\mathbf{n}+\mathbf{e}_{a,i})
\label{pi_eq1}
\end{equation}
and
\begin{eqnarray}
&&r_i\hat{\pi}(A|i,{\mathbf n})=n_{A,i}+\theta
+\frac{(1-b)\gamma'}{d}\sum_{j\neq i}\hat{\pi}(A|j,\mathbf{n})
\nonumber\\
&&
+\frac{2b\gamma'}{d}\sum_{j\neq i}\left[
\frac{(n_i+1)^2}{n_{A,i}+1}\hat{\pi}(A|i,\mathbf{n})
\hat{\pi}(a|j,\mathbf{n}+\mathbf{e}_{A,i})
+(n_i+1)\hat{\pi}(a|i,\mathbf{n})
\hat{\pi}(A|j,\mathbf{n}+\mathbf{e}_{a,i})\right],\nonumber\\
\label{pi_eq2}
\end{eqnarray}
where $r_i=n_i+2\theta+\gamma$. The system has the solution
\begin{equation}
\hat{\pi}(\alpha|i,\mathbf{n})
=\left(\frac{n_{\alpha,i}+\theta}{\gamma'/d}
+\sum_{j=1}^d\frac{n_{\alpha,j}-n_{\alpha,i}}{\gamma'/d+r_j}\right)
\left(
\frac{\gamma'/d+r_i}{\gamma'/d}-\sum_{j=1}^d\frac{\gamma'/d+r_i}{\gamma'/d+r_j}
\right)^{-1}
+O(b).\label{pi_sol}
\end{equation}
When $b=0$ (\ref{pi_sol}) is the exact probability that an additional gene
taken from deme $i$ is of allelic type $\alpha$ given that we
have a configuration $\mathbf n$. Even for $b>0$, solving quadratic system
of equations (\ref{pi_eq1}) and (\ref{pi_eq2}) is computationally
expensive. Using the expression of (\ref{pi_sol}) with $b=0$ and renormalizing
the proposal distribution is practically useful.

From Bayes' rule, $p(H_{k-1}|H_k)=p(H_k|H_{k-1})p(H_{k-1})/p(H_k)$, but 
$p(H_{k-1})/p(H_k)$ are unknown. The importance-sampling proposal 
distribution is obtained by substituting $\hat{\pi}$ for $\pi$ in 
$p(H_{k-1})/p(H_k)$, with the importance weights is given by
\begin{equation} 
\frac{p(H_k|H_{k-1})}{\hat{p}(H_{k-1}|H_k)}=
\frac{\hat{p}(H_k)}{\hat{p}(H_{k-1})}.
\end{equation}
For example, in the case $H_k=\mathbf n$ and
$H_{k-1}={\mathbf n}+\mathbf{e}_{a,i}-\mathbf{e}_{A,i}+\mathbf{e}_{A,j}$,
the proposal distribution is
\begin{eqnarray}
\hat{p}(H_{k-1}|H_k)
&=&p(H_k|H_{k-1})
\frac{\hat{p}({\mathbf n}+\mathbf{e}_{a,i}-\mathbf{e}_{A,i}+\mathbf{e}_{A,j})}
{\hat{p}(\mathbf{n}-\mathbf{e}_{A,i})}
\frac{\hat{p}(\mathbf{n}-\mathbf{e}_{A,i})}{\hat{p}(\mathbf{n})}\nonumber\\
&=&
\frac{2b\gamma'}{d}\frac{(n_{a,i}+1)(n_{A,j}+1)}{r(\mathbf{n})(n_j+1)}
\nonumber\\
&&\times
\frac
{n_i(n_j+1)}{(n_{a,i}+1)(n_{A,j}+1)}
\hat{\pi}(A|j,\mathbf{n}+\mathbf{e}_{a,i}-\mathbf{e}_{A,i})
\hat{\pi}(a|i,\mathbf{n}-\mathbf{e}_{A,i})\nonumber\\
&&\times
\frac{n_{A,i}}{n_i\hat{\pi}(A|i,\mathbf{n}-\mathbf{e}_{A,i})}\nonumber\\
&=&\frac{
2b\gamma n_{A,i}
\hat{\pi}(A|j,\mathbf{n}+\mathbf{e}_{a,i}-\mathbf{e}_{A,i})
\hat{\pi}(a|i,\mathbf{n}-\mathbf{e}_{A,i})}
{(n-1)r(\mathbf{n})\hat{\pi}(A|i,\mathbf{n}-\mathbf{e}_{A,i})},
\end{eqnarray}
and the importance weight is
\begin{equation}
\frac{
(n_{a,i}+1)(n_{A,j}+1)\hat{\pi}(A|i,\mathbf{n}-\mathbf{e}_{A,i})}
{n_{A,i}(n_j+1)
\hat{\pi}(A|j,\mathbf{n}+\mathbf{e}_{a,i}-\mathbf{e}_{A,i})
\hat{\pi}(a|i,\mathbf{n}-\mathbf{e}_{A,i})}.
\end{equation}
The proposal distribution and respective importance weights for the other 
cases are summarized in Table 1.

In (\ref{IS_sim}) $P(H_{-m})$ equals to either $\rho$ or $1-\rho$. 
A closed form expression for $\rho$ is not available, but it is easily
obtained by using the perfect simulation 
(coupling from the past)~\cite{ProppWilson1996,Fearnhead2001}.
For the perfect simulation, a parent independent mutation model, where 
an allele mutates to $A$ and $a$ with equal probabilities conditional 
on a mutation occurring with the mutation rate $2u$, is useful.
The parent independent mutation model is probabilistically equivalent 
to the mutation model discussed in previous sections, but the treatment
in the perfect simulation becomes much simpler. 

Consider simulating a sample of size one from the ancestral bias graph.
We can prove the following theorem on the expected number of events until 
the ancestral bias graph for a sample of size one couples.

\begin{theorem}
Consider the strong conversion limit ($dNb\rightarrow\beta$ as 
$N\rightarrow\infty$). Let $T$ be an exponential random variable with rate 
$\theta/4$. Let $M$ be the number of events in an ancestral bias graph 
initiated with a single branch, up to time $T$ in the past. The expected
number of events until the ancestral bias graph couples is bounded above 
by ${\mathbb E}[M]$. Furthermore,
\begin{equation}
\lim_{\beta\rightarrow\infty}{\mathbb E}
\left[\frac{M}{\beta^2}\right]\le\frac{(4c)^2}{\theta}.
\end{equation}
\end{theorem}
 
\begin{proof}
For the first part the proof is the same as that of Theorem~1 of 
\cite{Fearnhead2001}. For the second part, according to 
Remark~\ref{convASG} of Theorem~\ref{stlimit} the problem is reduced to 
find a bound for the expected number of events until the ancestral selection
graph with selection intensity $2\beta c/(1+c)$ initiated with a single branch
couples. Theorem~1 of \cite{Fearnhead2001} gives the required result.
\end{proof}

\section{Example: Mouse histone gene family}

Exon sequences of members of a mouse histone H2A gene family (single exon 
gene and 393 base pairs in length), which consists of 20 gene 
copies distributed at 6 unlinked loci in the mouse genome, were retrieved 
from Ensembl release 63 (http://www.ensembl.org/). The GC content at 
the third codon position was 90.2\%, which is significantly higher than 
average GC content in the mouse genome. The substitution rate was estimated 
to be $2.0\times 10^{-9}$ per site per generation, noting that the sequence 
divergence between the mouse sequence and the homologous rat sequence at 
the third codon position is 16\%, assuming the mouse-rat divergence time is 
20 million years and the average generation time is 0.5 years. If substitution
occurs symmetrically among nucleotides, 2/3 of substitutions occur between AT 
and GC. Under neutrality, the mutation rate between AT and GC is 
$u=1.3\times 10^{-9}$. The actual process of conversion involves a piece of 
a gene. Here, a nucleotide site was considered to be a unit of conversion and 
linkage among them was also ignored. Nucleotides of the third codon position 
were analyzed, where AT and GC nucleotides are regarded as allelic types $a$ 
and $A$, respectively. Assuming that the data set comes from $d=20$ loci of 
a single chromosome and the maximum likelihood estimate of $\theta$, $\gamma$,
and $b$, which jointly maximize the composite likelihood of the 131 
nucleotide sites was obtained by using the importance-sampling algorithm 
introduced in Section 7. The estimates were 
$\hat{\theta}=0.0008$, $\hat{\gamma}=0.04$ and $\hat{b}=0.07$. 
Then, the conversion rate per site per nucleotide would be 
$\hat{c}=6.5\times 10^{-8}$.

\noindent
{\bf Acknowledgments}

The author thanks Brian Charlesworth for suggesting the subject of this 
paper to him. He also thanks Robert Griffiths for comments on the importance
samping, Tokuzo Shiga for comments in connection with Sections 3 and 5, and 
Tomoko Ohta for comments on ectopic gene conversion.

\newpage

{\arrayrulewidth=.8pt
\renewcommand\arraystretch{1.6}
\begin{tabular}{ccc}
\hline
$H_{k-1}$&Proposal distribution&Importance weight\\
\hline
${\mathbf n}-\mathbf{e}_{a,i}$&
$\displaystyle
\frac{n_{a,i}(n_{a,i}-1)}
{r(\mathbf{n})\hat{\pi}(a|i,\mathbf{n}-\mathbf{e}_{a,i})}$&
$\displaystyle
\frac{n_i}{n_{a,i}}\hat{\pi}(a|i,\mathbf{n}-\mathbf{e}_{a,i})$\\
${\mathbf n}-\mathbf{e}_{A,i}$&
$\displaystyle
\frac{n_{A,i}(n_{A,i}-1)}
{r(\mathbf{n})\hat{\pi}(A|i,\mathbf{n}-\mathbf{e}_{A,i})}$&
$\displaystyle
\frac{n_i}{n_{A,i}}\hat{\pi}(A|i,\mathbf{n}-\mathbf{e}_{A,i})$\\
${\mathbf n}-\mathbf{e}_{a,i}+\mathbf{e}_{A,i}$&
$\displaystyle
\frac{\theta n_{a,i}\hat{\pi}(A|i,\mathbf{n}-\mathbf{e}_{a,i})}
{r(\mathbf{n})\hat{\pi}(a|i,\mathbf{n}-\mathbf{e}_{a,i})}$&
$\displaystyle
\frac{(n_{A,i}+1)\hat{\pi}(a|i,\mathbf{n}-\mathbf{e}_{a,i})}
{n_{a,i}\hat{\pi}(A|i,\mathbf{n}-\mathbf{e}_{a,i})}$\\
${\mathbf n}+\mathbf{e}_{a,i}-\mathbf{e}_{A,i}$&
$\displaystyle
\frac{\theta n_{A,i}\hat{\pi}(a|i,\mathbf{n}-\mathbf{e}_{A,i})}
{r(\mathbf{n})\hat{\pi}(A|i,\mathbf{n}-\mathbf{e}_{A,i})}$&
$\displaystyle
\frac{(n_{a,i}+1)\hat{\pi}(A|i,\mathbf{n}-\mathbf{e}_{A,i})}
{n_{A,i}\hat{\pi}(a|i,\mathbf{n}-\mathbf{e}_{A,i})}$\\
${\mathbf n}-\mathbf{e}_{a,i}+\mathbf{e}_{a,j}$&
$\displaystyle
\frac{(1-b)\gamma' n_{a,i}\hat{\pi}(a|j,\mathbf{n}-\mathbf{e}_{a,i})}
{dr(\mathbf{n})\hat{\pi}(a|i,\mathbf{n}-\mathbf{e}_{a,i})}$&
$\displaystyle
\frac{n_i(n_{a,i}+1)\hat{\pi}(a|i,\mathbf{n}-\mathbf{e}_{a,i})}
{n_{a,i}(n_j+1)\hat{\pi}(a|j,\mathbf{n}-\mathbf{e}_{a,i})}$\\
${\mathbf n}-\mathbf{e}_{A,i}+\mathbf{e}_{A,j}$&
$\displaystyle
\frac{(1-b)\gamma' n_{A,i}\hat{\pi}(A|j,\mathbf{n}-\mathbf{e}_{A,i})}
{dr(\mathbf{n})\hat{\pi}(A|i,\mathbf{n}-\mathbf{e}_{A,i})}$&
$\displaystyle
\frac{n_i(n_{A,j}+1)\hat{\pi}(A|i,\mathbf{n}-\mathbf{e}_{A,i})}
{n_{A,i}(n_j+1)\hat{\pi}(A|j,\mathbf{n}-\mathbf{e}_{A,i})}$\\
${\mathbf n}+\mathbf{e}_{a,j}$&
$\displaystyle
\frac{2b\gamma' n_i\hat{\pi}(a|j,\mathbf{n})}
{dr(\mathbf{n})}$&
$\displaystyle
\frac{n_{a,j}+1}
{(n_j+1)\hat{\pi}(a|j,\mathbf{n})}$\\
${\mathbf n}+\mathbf{e}_{A,j}$&
$\displaystyle
\frac{2b\gamma' n_{A,i}\hat{\pi}(A|j,\mathbf{n})}
{dr(\mathbf{n})}$&
$\displaystyle
\frac{n_{A,j}+1}
{(n_j+1)\hat{\pi}(A|j,\mathbf{n})}$\\
\hline
\end{tabular}}

\begin{figure}
\includegraphics[width=\textwidth]{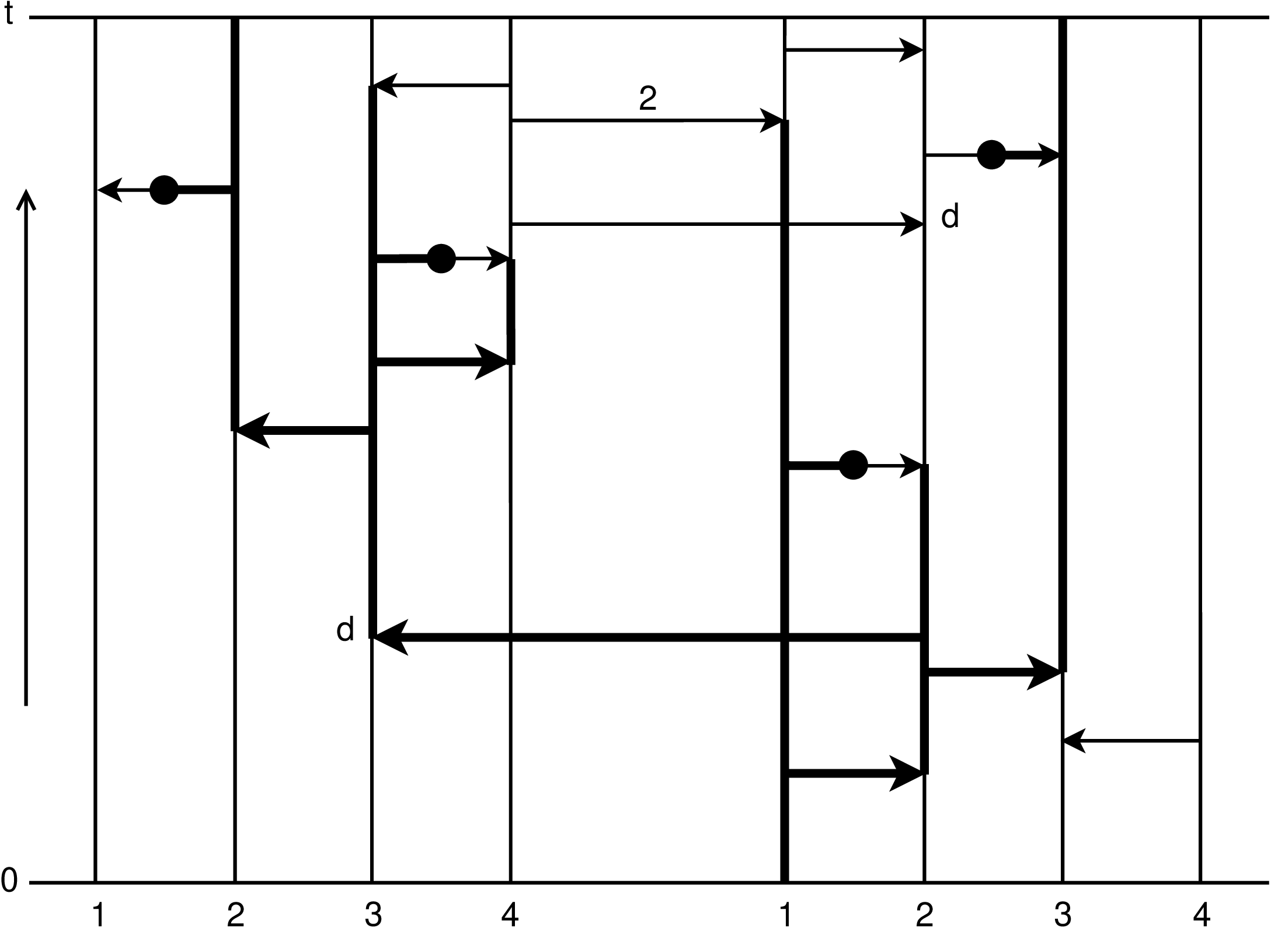}

\caption{A graphical representation for the biased voter model for 
the case $d=2$ and $N=4$. If initially the set of $a$'s is $\{1\}\in I_2$, 
then at time $t$, the set of $a$'s is $\{2\}\in I_1$ and $\{3\}\in I_2$. 
The paths of $a$'s are indicated by thick lines.}
\end{figure}

\begin{figure}
\includegraphics[width=\textwidth]{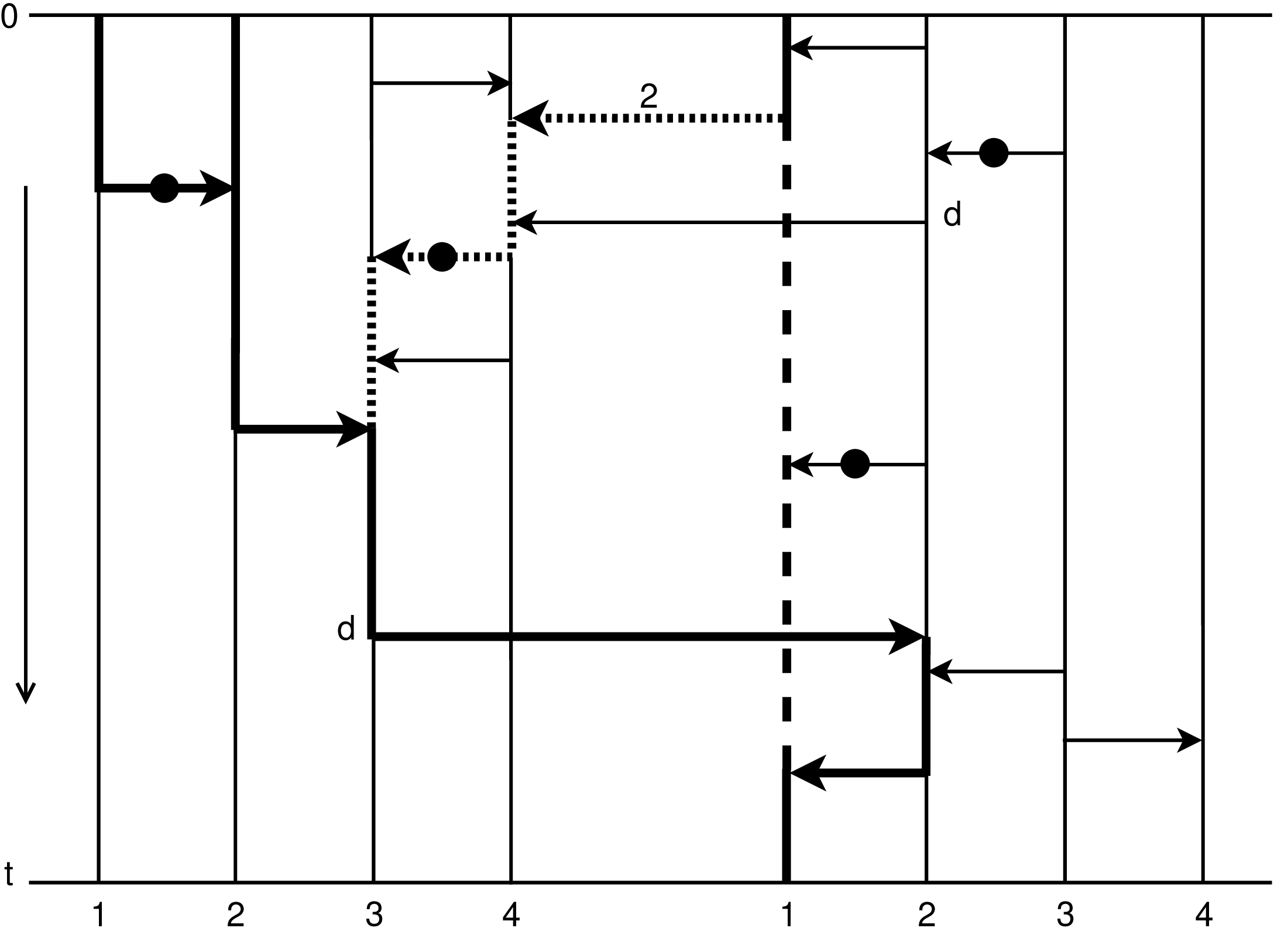}
\caption{A graphical representation for the dual process of the biased
voter model. The ancestral history of a sample, consists of 
individuals at sites $\{1,2\}\in I_1$ and $\{1\}\in I_2$ at dual time 0 
is indicated by thick lines. The ultimate ancestor is in $\{1\}\in I_2$,
at dual time $t$. If the ultimate ancestor is $a$, then the true genealogy 
contains the dotted line and does not contain the dashed line.}
\end{figure}


\begin{thebibliography}{99}

\footnotesize

\bibitem[Berglund et al., 2009]{Berglund2009}
{\sc Berglund, J, Pollard, K.~S. and Webster, M.~T.} (2009).
Hotspots of biased nucleotide substitutions in human genes. 
{\em PLoS Biol.} {\bf 7,} e1000026.

\bibitem[De Iorio and Griffiths, 2004a]{DeIorioGriffiths2004a}
{\sc De Iorio, M. and Griffiths, R.~C.} (2004).
Importance sampling on coalescent histories. I.
{\em Adv. Appl. Prob.} {\bf 36,} 417--433.

\bibitem[De Iorio and Griffiths, 2004b]{DeIorioGriffiths2004b}
{\sc De Iorio, M. and Griffiths, R.~C.} (2004).
Importance sampling on coalescent histories. II: subdivided population 
models.
{\em Adv. Appl. Prob.} {\bf 36,} 434--454.

\bibitem[Donnelly, 1984]{Donnelly1984}
{\sc Donnelly, P.} (1984).
The transient behavior of the Moran model in population genetics.
{\em Proc. Camb. Phil. Soc.} {\bf 95,} 349--358.

\bibitem[Ethier and Nagylaki, 1980]{EthierNagylaki1980}
{\sc Ethier, S.~N. and Nagylaki, T.} (1980).
Diffusion approximations of Markov chains with two time scales and 
applications to population genetics.
{\em Adv. Appl. Prob.} {\bf 12,} 14--49.

\bibitem[Fearnhead, 2001]{Fearnhead2001}
{\sc Fearnhead, P.} (2001).
Perfect simulation from population genetic models with selection.
{\em Theor. Popul. Biol.} {\bf 59,} 263--279.

\bibitem[Galtier, 2003]{Galtier2003}
{\sc Galtier, N.} (2003).
Gene conversion drives GC content evolution in mammalian histones.
{\em Trends in Genet.} {\bf 19,} 65--68.

\bibitem[Griffiths and Tavar\'{e}, 1994]{GriffithsTavare1994}
{\sc Griffiths, R.~C. and Tavar\'{e}, S.} (1994).
Simulating probability distribution in the coalescent.
{\em Theor. Popul. Biol.} {\bf 46,} 131--159.


\bibitem[Harris, 1972]{Harris1972}
{\sc Harris, T.~E.} (1972).
Nearest neighbor Markov interaction process on multidimensional lattices.
{\em Adv. Math.} {\bf 9,} 66--89.

\bibitem[Harris, 1976]{Harris1976}
{\sc Harris, T.~E.} (1976).
On a class of set-values Markov process.
{\em Ann. Probab.} {\bf 4,} 175--194.

\bibitem[Kingman, 1982]{Kingman1982}
{\sc Kingman, J.~F.~C.} (1982).
The coalescent.
{\em Stoch. Proc. Appl.} {\bf 13,} 235--248.

\bibitem[Krone and Neuhauser, 1997]{KroneNeuhauser1997}
{\sc Krone, S.~M. and Neuhauser, C.} (1997).
Ancestral process with selection. 
{\em Theor. Popul. Biol.} {\bf 51,} 210--237.


\bibitem[Mano, 2009]{Mano2009}
{\sc Mano, S.} (2009).
Duality, ancestral and diffusion processes in models with selection.
{\em Theor. Popul. Biol.} {\bf 75,} 164--175.

\bibitem[Mano and Innan, 2008]{ManoInnan2008}
{\sc Mano, S. and Innan, H.} (2008).
The evolutionary rate of duplicated genes under concerted evolution.
{\em Genetics} {\bf 181,} 493--505.

\bibitem[Nagylaki, 1980]{Nagylaki1980}
{\sc Nagylaki, T.} (1980).
The strong migration limit in geographically structured populations.
{\em J. Math. Biol.} {\bf 9,} 101--114.

\bibitem[Nagylaki, 1983]{Nagylaki1983}
{\sc Nagylaki, T.} (1983). Evolution of a finite population under
gene conversion. {\em Proc. Natl. Acad. Sci. USA} {\bf 80,} 6278--6281.

\bibitem[Nagylaki and Petes, 1982]{NagylakiPetes1982}
{\sc Nagylaki, T. and Petes, T.} (1982).
Interchromosomal gene conversion and the maintenance of sequence homogeneity
among repeated genes.
{\em Genetics} {\bf 100,} 315--337.

\bibitem[Ohta, 1982]{Ohta1982}
{\sc Ohta, T.} (1982).
Allelic and nonallelic homology of a supergene family.
{\em Proc. Natl. Acad. Sci. USA} {\bf 79,} 3251--3254. 


\bibitem[Propp and Wilson, 1996]{ProppWilson1996}
{\sc Propp, J.~G. and Wilson, D.~B.} (1996).
Exact sampling with coupled Markov chains and applications to statistical
mechanics.
{\em Random Struct. Algorithms} {\bf 9,} 223--252.

\bibitem[Shiga and Uchiyama, 1986]{ShigaUchiyama1986}
{\sc Shiga, T. and Uchiyama, K.} (1986).
Stationary state and their stability of the stepping stone model involving
mutation and selection.
{\em Prob. Theor. Rel. Fields} {\bf 73,} 87--117.

\bibitem[Stephens and Donnelly, 2000]{StephensDonnelly2000}
{\sc Stephens, M. and Donnelly, P.} (2000).
Inference in molecular population genetics.
{\em J. R. Statist. Soc.} {\bf B 62,} 605--655.

\bibitem[Walsh, 1985]{Walsh1985}
{\sc Walsh, J.~B.} (1985).
Interaction of selection and biased gene conversion in a multigene family.
{\em Proc. Natl. Acad. Sci. USA} {\bf 82,} 153--157. 

\bibitem[Wright, 1951]{Wright1951}
{\sc Wright, S.} (1951).
The genetical structure of populations.
{\em Ann. Eugen.} {\bf 15,} 323--354.


\end{thebibliography}
\end{document}